\documentclass[a4paper]{article}
\usepackage[preprint, nonatbib]{neurips_2020}
\usepackage[utf8]{inputenc} 
\usepackage[T1]{fontenc}    
\usepackage{hyperref}       
\usepackage{url}            
\usepackage{booktabs}       
\usepackage{amsfonts}       
\usepackage{nicefrac}       
\usepackage{microtype}      
\usepackage{caption}
\usepackage[english]{babel}
\usepackage[ruled,vlined]{algorithm2e}

\usepackage{amsmath}
\usepackage{graphicx}
\usepackage[colorinlistoftodos]{todonotes}

\usepackage{subcaption}
\usepackage{url}

\title{Is a trillion trees enough?}

\author{

C\'edric Mesnage \texttt{cedric@universalresearch.center}\\
\href{http://www.universalresearch.center}{Universal Research Center}

\And

Michalis Vlachos \texttt{Michalis.Vlachos@unil.ch} \\
\href{http://www.unil.ch}{University of Lausanne}
}

\begin{document}

\maketitle

\begin{abstract}

  Global warming is a major environmental concern of our times,  \cite{IPCC_2018}. It has been suggested that the planting of trees could constitute a way of mitigating the adverse effects of the increasing anthropogenic carbon emissions. We developed a simple data-driven model to predict the  global average temperature in diverse scenarios to better understand the outcome of human interventions. We find that to remain under the 1.5 degree of anomaly set by the Paris agreement, the fossil-fuel energy production needs to decrease to at least 10\% of the projected figure in 2050 combined with planting a trillion trees.
  
\end{abstract}

\section{Introduction}
No, it is not enough. Our analysis shows that planting trees is not sufficient, as a singular measure, to effectively slow down the increase in global temperature. The gas emissions due  to the increased industrialization and  the  increase  of  the population have invariably led to a rise of the global  average  temperature  and correspondingly to a rise in ocean temperatures and the sea  level.  It has been suggested by many scientific panels that this increase in the global average temperature may be the cause for the extinction of various species as well as other natural disasters \cite{van2006impacts}. For example, recently it has been shown that the largest king-penguin colony has experienced a drastic reduction in its population (see \cite{weimerskirch2018massive}), reducing from  2  million  individuals  to 90,000. Another alarming fact is the change in the ocean microbiotic life. \cite{Moran20150371} have shown that bacteria are becoming smaller and there are more of them as a result of global warming. What would be the effect of this for ocean life in general? One way to stop the increase of the global temperature would be to reduce gas emissions  \cite{smith2019current} with the goal of having the temperature rise not more than  1.5  degrees Celsius, as recommended by the Paris agreement \cite{change2018paris}. 

It has been suggested that planting of new trees could offer a way of capturing back the newly created CO2 emissions, thus neutralizing the effects of increased industrialization.  Currently, there  are  approximately 3 trillion trees on the planet, with 6 trillion at the dawn of humanity~\cite{crowther2015mapping}. Another 1.2 trillion would be necessary to absorb the excess of CO2 in the atmosphere according to \cite{goymer2018trillion}. Identifying major industries and activities which contribute to gas emissions and convincing people to find alternatives is essential as well as reforestation efforts and planting trees in arid areas.

In this paper, we are examining scenarios of whether planting a trillion trees could constitute a solution to neutralizing the increasing CO2 emissions. We understand this is a controversial topic and it is not expected to be the proper solution to the global warming problem. Nonetheless, such an approach can sequester carbon dioxide from the atmosphere. This paper showcases some preliminary evaluations in understanding the situation and trying to predict the effect of diverse scenarios on the global average temperature. We find that combining planting trees and bringing energy production to zero carbon would keep the global average temperature anomaly below 1.5 degrees Celcius. We discuss limitations and assumptions of our approach, which understandably only considers a limited number of factors in the overall complex environmental ecosystem.

\section{Relevance to Machine Learning}

Following the vision set by \cite{rolnick2019tackling} we follow a simple machine learning methodology to understand how to positively influence climate change. Our approach is data-driven, also supported by simple machine learning techniques and considering multiple variables for predicting the outcome of various `what if' scenarios. We use multivariate polynomial ridge regression \cite{haitovsky1987multivariate} of degree 2 to train our model based on the global average temperature and the amount of CO2 in the atmosphere. We use this model to make predictions under various scenarios by changing the values of the CO2 for the upcoming 30 years. We use the amount of CO2 sequestrated (absorbed) by trees as a proxy for inferring the reduction in the CO2 levels. 
To predict the trend in future CO2 levels and temperatures we use historical data offered by the National Oceanic and Atmospheric Administration (NOAA) as well as NASA \footnote{\url{https://www.ncdc.noaa.gov/cag/global/time-series/globe/land_ocean/ytd/12/1880-2019.csv}}\footnote{\url{ftp://aftp.cmdl.noaa.gov/products/trends/co2/co2_mm_mlo.txt}}\footnote{\url{https://data.giss.nasa.gov/modelforce/ghgases/Fig1A.ext.txt}}.





\section{Scenarios}

We analyse various scenarios of possible evolutions in CO2 and temperature levels from 2020 to 2050. The code to produce our predictions and scenarios is available on Google Colab \footnote{\url{https://colab.research.google.com/drive/1mAIXSHnt5z4f9C5e7y6EqNHeuvVr-N7b}}.

\subsection{How much carbon?}

The potential amount of absorbed carbon dioxide by planting trees calculated by \cite{Veldmaneaay7976}, is much lower than the 205 gigatons envisioned by \cite{crowther2015mapping}. Between 42 and 107 gigatons depending on the effect of the change of albedo, increased fires or other disasters due to the increase of canopy cover. We consider three main scenarios, spanning from no reduction of gas emissions, to a small reduction, to zero-carbon.  For each scenario we examine what would transpire if we a) did not plant any new trees (and assume the current amount of trees remains constant), b) plant a trillion trees and consider 42 GtC would be sequestered, c) plant a trillion trees and consider 107 GtC would be sequestered.

\subsection{Scenario 1:  A continuous increase of gas emissions by 2050}

First, we consider the scenario (a) in which no action is taken and the CO2 levels in the atmosphere continue to rise. We use a quadratic regression to predict the rise of CO2 and use this prediction as input to predict the rise of the temperature in our multivariate model. In this case our model predicts the temperature will rise to 2.17 degrees anomaly above the pre-industrial levels,  shown in Figure \ref{fig:plot1a}.

We consider the previous prediction for the rise of CO2 levels and deduce from it a logarithmically spaced vector of 0 to 2 ppm (b) over the next 30 years summing up to 42GtC or to 15 ppm summing to 107 GtC (c). This constructed CO2 levels are used to predict the global average temperature. Giving the scenario shown in Figure \ref{fig:plot1b} and \ref{fig:plot1c}. We observe a decrease in the global average temperature which would remain under the 2 degree boundary of the Paris agreement at 1.95 degrees Celsius.

\subsection{Scenario 2: A realistic decrease of gas emissions by 2050}

The oil and gas companies predict an increase of 25\% of non-fossil energy as a part of the total energy production in the next 30 years \cite{bakken2020}. Emissions would keep on rising but at a decreased rate of 100\% to 69\% over the next 30 years. We apply this rate in equation \ref{eq:2} to compute the CO2 levels of this scenario and deduce the potential absorption for each year in the case of the prime scenario. The CO2 level for a certain year is defined by the CO2 level of the previous year ($CO2_{year-1}$) to which we add the predicted increase times the reduction rate for that year.

This scenario is more realistic than the following one of aiming for zero-carbon emissions by 2050. If no carbon sequestration is put in place the temperature would reach a 1.87 degree anomaly, in Figure \ref{fig:plot2a}. With the absorption thanks to trees, the temperature would reach 1.67 degrees, in Figure \ref{fig:plot2c}.

\begin{small}
\begin{equation}
CO2_{year} = CO2_{year-1} + (CO2\_pred[year] - CO2\_pred[year-1]) * rate[year] - absorption[year]\label{eq:2}
\end{equation}
\end{small}

\subsection{Scenario 3:  Zero-carbon by 2050 with/without a trillion trees planted}

The scenario of reaching zero carbon by 2050 is implemented by applying a linear reduction rate from 100\% to 0\% over the next 30 years to the CO2 increase predicted originally, we use this rate in equation \ref{eq:2}. In this case, the temperature remains below 1.5 degrees anomaly (Figure \ref{fig:plot3a}) at 1.28 and decreases to 1.12 degrees when combined with planting trees, in Figure \ref{fig:plot3c}.




\subsection{Scenario 4: How much do we have to decrease gas emissions to remain under 1.5 degrees}

We develop a simple algorithm to find the minimum decrease of gas emissions required such that the global average temperature remains under 1.5 degrees when combined with planting trees to absorb 107 GtC. The least decrease of gas emissions required is to 10\% by 2050.

{\centering
\begin{table}[ht]
\centering
\begin{tabular}{c|c|c|c}
 & No trees & 42 GtC & 107 GtC \\
\parbox[c]{7pt}{\rotatebox{90}{Continuous increase}} &
\begin{subfigure}{.3\textwidth}\centering\includegraphics[width=\columnwidth]{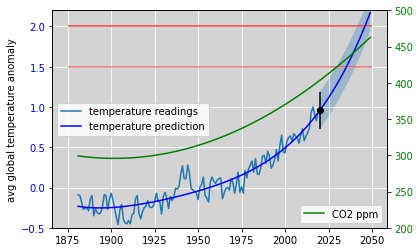}\caption{Scenario 1 a}\label{fig:plot1a}\end{subfigure}& \begin{subfigure}{.3\textwidth}\centering\includegraphics[width=\columnwidth]{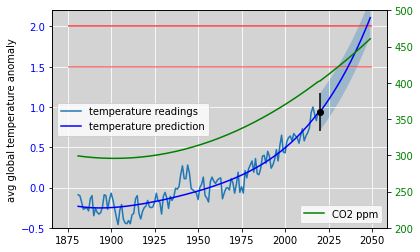}\caption{Scenario 1 b}\label{fig:plot1b}\end{subfigure}
& \begin{subfigure}{.3\textwidth}\centering\includegraphics[width=\columnwidth]{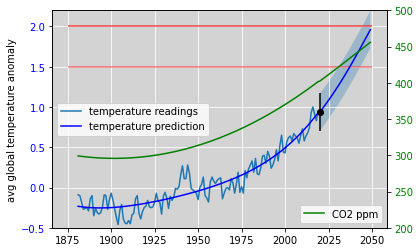}\caption{Scenario 1 c}\label{fig:plot1c}\end{subfigure}
 \\
\hline
\parbox[c]{7pt}{\rotatebox{90}{Decrease to 69\%}}&
\begin{subfigure}{.3\textwidth}\centering\includegraphics[width=\columnwidth]{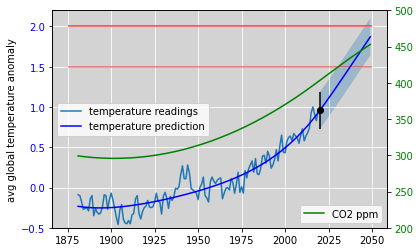}\caption{Scenario 2 a}\label{fig:plot2a}\end{subfigure}& \begin{subfigure}{.3\textwidth}\centering\includegraphics[width=\columnwidth]{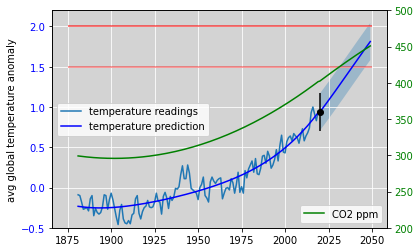}\caption{Scenario 2 b}\label{fig:plot2b}\end{subfigure}
& \begin{subfigure}{.3\textwidth}\centering\includegraphics[width=\columnwidth]{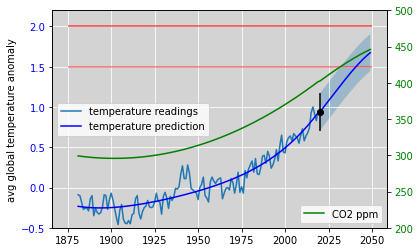}\caption{Scenario 2 c}\label{fig:plot2c}\end{subfigure} \\ 
\hline
\parbox[c]{7pt}{\rotatebox{90}{Zero-carbon}}&
\begin{subfigure}{.3\textwidth}\centering\includegraphics[width=\columnwidth]{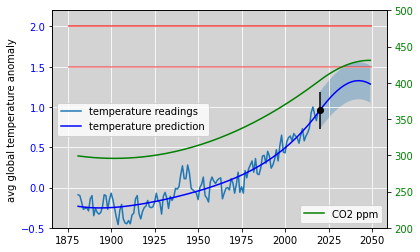}\caption{Scenario 3 a}\label{fig:plot3a}\end{subfigure}& \begin{subfigure}{.3\textwidth}\centering\includegraphics[width=\columnwidth]{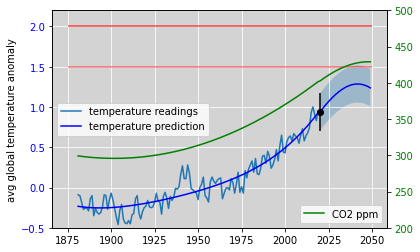}\caption{Scenario 3 b}\label{fig:plot3b}\end{subfigure}
& \begin{subfigure}{.3\textwidth}\centering\includegraphics[width=\columnwidth]{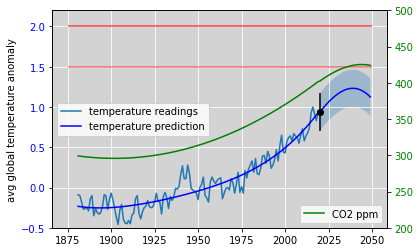}\caption{Scenario 3 c}\label{fig:plot3c}\end{subfigure} \\
\hline
\parbox[c]{7pt}{\rotatebox{90}{Decrease to 10\%}}&
\begin{subfigure}{.3\textwidth}\centering\includegraphics[width=\columnwidth]{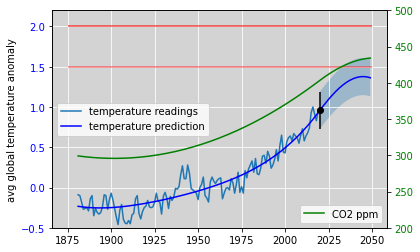}\caption{Scenario 4 a}\label{fig:plot4a}\end{subfigure}& \begin{subfigure}{.3\textwidth}\centering\includegraphics[width=\columnwidth]{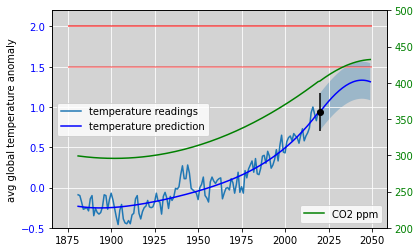}\caption{Scenario 4 b}\label{fig:plot4b}\end{subfigure}
& \begin{subfigure}{.3\textwidth}\centering\includegraphics[width=\columnwidth]{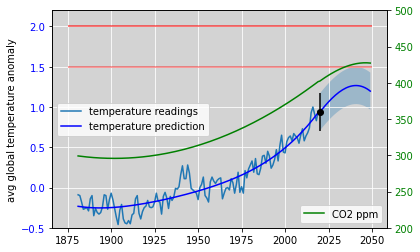}\caption{Scenario 4 c}\label{fig:plot4c}\end{subfigure} \\
\end{tabular}
\caption{Table of prediction plots for each scenario. The columns represent whether we plant trees or not; the rows represent the three scenarios on the change in gas emissions. The light blue area shows the 95\% confidence interval as two times the root mean squared error (0.114).}
\label{tab:mytable}
\end{table}
}





\section{Discussion and assumptions}

Our analysis assumes a connection between CO2 levels and the global average temperature. We know that a correlation exists but this does not imply a direct causal relationship between the two variables. There is still debate whether a reduction of the CO2 in the atmosphere would result directly in a decrease of the temperature as our models show, in \cite{pietrafesa2019global} Pietrafesa et al. use Granger causality to highlight the relation between carbon emissions and global average temperature; in \cite{sippel2020climate} Sippel et al. analyse the relation between various factors and temperature and conclude the increase is due to human factors. There are many other factors involved when planting trees; the albedo caused by the change of surface, i.e. when the trees are grown the land surface becomes darker compared to having ice or sand which reflects sunlight, dark surface absorbs light and heat; the species of trees being planted, different varieties will absorb CO2 at different rates, mangrove, oaks and chestnut seem to be good candidates although species to be planted depend on the local climate \cite{ramachandran2009agroforestry}.

In our study, we assumed that the capacity of the Earth's oceans to absorb CO2 will remain constant. However, there have been raising concerns that the increase in the average ocean temperatures and acidity may affect the capacity of the ocean's algae to continue absorbing CO2 at the same rate \cite{doney2006dangers}.

Reforestation and afforestation efforts as well as the development of a climate economy \cite{dao2019gainforest} to protect existing forests are essential to solve the climate crisis.

\subsection{Climate models}

Why not use existing climate models? In this paper we developed our own model in order to have more control on changing parameters, such as the amount of potential carbon being sequestered by trees and its speed, as well as reduction of gas emissions. This enables us to better understand the relation between these variables and the issues which arise with modeling. We are now convinced with the results. In comparison, using a climate model approach, \cite{enroads} enables to modify parameters by moving sliders; we find it difficult to understand from a data science perspective, how the model is constructed and what are the parameters values being used. Nevertheless our findings are similar to the findings given by climate models. This corroborates both climate models and our model.

\subsection{Further research}

There is a large margin of uncertainty when it comes to evaluate how much carbon can be sequestered by planting trees, i.e. between 42 GtC and 107 GtC. As \cite{Veldmaneaay7976} consider not to include efforts in reforestation in biomes with an important change in albedo such as boreal forests or the tundra ; neither tropical and temperate grasslands where there is an increased risk of fire and a potential decrease of biodiversity. We think that completely removing these biomes which leads to a reduction from 107GtC to 42GtC of potential carbon sequestered is inexact and requires more research , for instance in predicting and measuring the effect of the change of albedo due to planting trees or planning where and how to plant trees to reduce risks of wild fires. These are areas where machine learning can help in order to get a more accurate picture of what reforestation can achieve.


\section{Conclusions}
While there exist several simulation-driven approaches to understand global warming, our approach is purely data-driven. Our model showed that without carbon sequestration involving planting trees no scenario remain under 1.5 degrees of anomaly (including the prediction error). We find that to remain under the 1.5 degree of anomaly set by the Paris agreement, the fossil-fuel energy production needs to decrease to 10\% of the projected figure combined with planting a trillion trees: thus the renewable part of energy production needs to progress from 19\% in 2020 to 91.9\% in 2050. In our view, both efforts of reforestation and energy transformation are necessary to solve climate change, confirming \cite{holl2020}. 
Researchers should see the present effort as a call for further data-driven approaches that will increase our understanding of this complex problem and our capacity in making rational decisions.





\bibliographystyle{abbrv}
\bibliography{bibliography}

\section*{Appendix : Incidence on energy production}


\begin{figure}[ht]
    \centering
    \includegraphics[width=.5\textwidth]{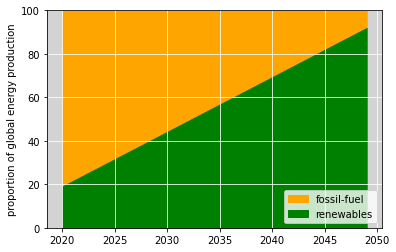}
        \captionsetup{labelformat=empty}
    \caption{Projected required evolution of the global energy production to remain under 1.5 degrees of anomaly with 107 gigatons of CO2 sequestered by trees by 2050. The renewable part of energy production needs to progress from 19\% in 2020 to 91.9\% in 2050.}
\end{figure}

\end{document}